\documentstyle[11pt]{article}

\def\emline#1#2#3#4#5#6{%
       \put(#1,#2){\special{em:moveto}}%
       \put(#4,#5){\special{em:lineto}}}
\def\newpic#1{}

\title{
\begin{flushright}
{\normalsize Yaroslavl State University\\
             Preprint YARU-HE-94/08\\
             hep-ph/9409450} \\[3cm]
\end{flushright}
       Rare Electroweak Processes
       $K^0_L \to \mu^+ \mu^-$ and $K^0_L \to \gamma \gamma$
       and Heavy Top Quark}

\author{Gvozdev A.A., Mikheev N.V. and Vassilevskaya L.A.\\
        Yaroslavl State University, Yaroslavl, Russia}

\date{}

\begin{document}

\maketitle

\begin{abstract}
A brief overview of the recent measurements of the branching ratio
of the rare $K_L^0 \to \mu^+ \mu^-$ decay in the context of their
agreement with the Standard Model (SM) is given. It is shown that
KEK result well correlates with the SM and B-physics, whereas the
BNL results are in conflict with the SM with the heavy top quark.
\end{abstract}

\vglue 3cm

\begin{center}
{\it Talk given at the 5th Conference on the Intersections\\
of Particle and Nuclear Physics,\\
St. Petersburg, Florida, USA, May 31 - June 6, 1994}
\end{center}

\newpage

For a long time the rare electroweak decays $K^0_L \to \mu^+ \mu^-$ and
$K^0_L \to \gamma \gamma$ come to the attention of
physicists. The connection between absorptive width of
$K^0_L \to \mu^+ \mu^-$ decay  calculated from imagine part of an
amplitude and total $K^0_L \to \gamma \gamma$ decay width follow from
the known unitarity relation:

\begin{equation}
Br_{abs} ( K^0_L \to \mu^+ \mu^- ) \simeq 1.2 \cdot 10^{-5} \,
Br ( K^0_L \to \gamma \gamma ) = (6.8 \pm 0.3) \cdot 10^{-9} .
\label{eq:ur}
\end{equation}

\noindent Here we used the experimental value of
$Br ( K^0_L \to \gamma \gamma ) = (5.70 \pm 0.27) \cdot 10^{-4}$.
This minimal value allowed by theory is known  as the
unitarity limit. It is curiously to emphasize that the amplitude of
$K^0_L \to \mu^+ \mu^-$ decay consists of two comparable contributions.
The first one is due to long distances ($r \sim 1/m_K$), where the
light quark contribution is essential, and the other come from
short distances ($r \le 1/m_W$), where the contributions of heavy
quarks dominate. At the first time this fact was shown  in the paper
of M.~Voloshin and E.~Shabalin~\cite{VSh}, where the estimation of the
$c$-quark mass from the width of the mentioned above decay
in the framework of the SM with two generations was
obtained. As time passed, it became aware the importance of
$K^0_L \to \mu^+ \mu^-$ decay as  one more source for the top quark
mass estimation (see, for example, Ref.~\cite{GN, B}).
In our papers~\cite{G1, G2, G3} the total amplitude of this process has
been calculated within the quark model approach.

In the quark approach  $K^0_L \to \mu^+ \mu^-$ amplitude is a sum of
one-loop ($1L$) and two-loop ($2L$) contributions. The first one
(through $W$ and $Z$) due to short distances $\sim 1/m_W$ where top
quark contribution dominates. As for the $2L$ contribution with two
photon intermediate state, medium ($\sim 1/m_c$)  and rather long
($\ge 1/m_K$) distances are essential. The evidence of the top quark
production at the $p \bar p$ collider (CDF Collab., FNAL)~\cite{CDF}
and also a further precision of the estimations of $V_{u b}$ and $V_{c b}$
CKM matrix elements improved by ARGUS and CLEO~\cite{Ali} allow us
to overview the modern status of $K^0_L \to \mu^+ \mu^-$ decay.
Namely, calculating the total decay amplitude in the framework
of the SM one can obtain the restriction on the decay width and,
by this means, investigate the agreement of the recent experimental
data with the SM.

We pointed out the principal importance of the relative sign between
$1L$ and $2L$ contributions.
Let us note that in the terms of the bare quarks the total decay
amplitude contains these contributions with opposite signs~\cite{VSh}.
The obtained by us amplitude~\cite{G1} in the limit $m_u^2 \ll m_K^2/4$,
$m_c^2 \gg m_K^2/4$ for the current $u$- and $c$-quarks agrees with
the result of Voloshin and Shabalin~\cite{VSh}.
However we emphasize that it is necessary to account the QCD corrections
to the
effective four-quark vertex in order to obtain a realistic result for $2L$
contribution. To that end, we used the renormgroup method by Vainstein,
Zakharov and Shifman~\cite{VZSh} for the mass scale $\mu$ down to
the typical hadronic scale $\mu_0 \simeq 2 \Lambda$
($\alpha_{st}(\mu_0) = 1$).
We have developed also a fenomenological method of the estimation of the
QCD corrections on the small scale interval $\mu_0 \le \mu \le m_K$~\cite{G2}.
To test the reliability of our method, we
calculated $\Gamma(K_L^0 \to e^+ e^- \gamma) /
\Gamma(K_L^0 \to \mu^+ \mu^- \gamma)$~\cite{G3} and showed that our result
is closed to one obtained within
the phenomenological pole model~\cite{B}.
Certainly we do not pretend to obtain an integral accuracy better than
$30 \div 40 \%$ in the description of the contributions of relatively
long distances ($r \le 1/ \mu_0$). However, the sign between $1L$ and
the real part of $2L$ contributions is fixed sufficiently reliable by
this way. Our main result is that the real part of $2L$ contribution
changes the sign if the QCD corrections take into account.
The change of the sign is connected with the behaviour of the integral
over the $u$-quark loop scale. This integral involves multiplicatively
the QCD formfactor of $(V-A)$ four-quark vertex which becomes
sufficiently large (more than unit in modulus) and negative
number on the interval $2 \Lambda \le \mu \le m_K$~\cite{VZSh}.

The expression for the total $K_L^0 \to \mu^+ \mu^-$
amplitude obtained by this way
has the form~\cite{G2}:

\begin{eqnarray}
{\cal M} ( K_L^0 \to \mu^+ \mu^- ) & \simeq & -  10^{-3} \, {\cal N} \,
\big \lbrace + [ ( 5.6 \pm 2.0 ) - i (44.7 \pm 0.9) ] \nonumber \\
&& + 2 + 10^3 \; \frac{F(m_t^2 / m_W^2)}{\sin^2 \theta_W} \;
\frac{\Re (V^*_{t d} V_{t s})}{\Re (V^*_{c d} V_{c s})} \big \rbrace ,
\label{eq:ampl} \\
F(x) & = & \frac{x}{4} \, \left [ \frac{4 - x}{1 - x} +
\frac{3 x \ln x}{(1 - x)^2} \right ] ,  \nonumber
\end{eqnarray}

\noindent where ${\cal N} = (\alpha / 4 \pi) \, G_F \, F_K \, m_\mu
\sin 2 \theta_C (\bar \mu \gamma_5 \mu)$, $F(x)$ is the well-known
function~\cite{MP}, $F_K$ is the formfactor of the $K$-meson, $m_\mu$ is
the muon mass, $\theta_C$ and $\theta_W$ are the Cabibbo and Weinberg
angles respectively. The first term in the curly braces describes
the $2L$ contribution.
We pretend only on the calculation of the real part of $2L$ contribution,
and take the imagine part from the unitarity relation~(\ref{eq:ur}).
The second and third terms of the amplitude~(\ref{eq:ampl}) describe the $c$-
and $t$-quark contributions respectively.

It should be noted that our expression~(\ref{eq:ampl}) for the total
decay amplitude is in contradiction with the result by Ko~\cite{Ko1}
in which the relative sign between the first and others terms of the
amplitude is negative, whereas in our expression it is the same.
The method developed by Ko~\cite{Ko1,Ko2} has disadvantage.
Namely, in these papers the dependence of the meson vertex
formfactors (for example, $\pi V V$) on the meson loop scale
was neglected.

To obtain the restriction on $K_L^0 \to \mu^+ \mu^-$ decay width,
we used the resent experimental data on the top quark mass~\cite{CDF}

\begin{displaymath}
m_t = 174 \pm 10^{+13}_{-12} = 174 \pm 16 \, GeV
\end{displaymath}

\noindent and on the parameters of CKM matrix
in the Wolfenstain representation~\cite{Ali}

\begin{eqnarray}
&& \lambda = \sin \theta_C \simeq 0.22 , \qquad A = 0.86 \pm 0.10,
\nonumber \\
&& \sqrt{\rho^2 + \eta^2} = 0.36 \pm 0.09 \qquad \mbox{which gives} \qquad
(1 - \rho) \ge 0.64 \pm 0.09 .
\nonumber
\end{eqnarray}

\noindent From Eq.~(\ref{eq:ampl}) the lower limit
on the $K_L^0 \to \mu^+ \mu^-$ decay width is following:

\begin{eqnarray}
&& \Delta Br (K_L^0 \to \mu^+ \mu^-) \cdot 10^9 =
\big [ Br (K_L^0 \to \mu^+ \mu^-) \cdot 10^9 - 6.8 \big ] \ge
\nonumber \\
&& 0.95 (1 \pm 0.13 \pm 0.08 \pm 0.08 \pm 0.04 \pm 0.12) = 0.95 (1 \pm 0.2)
\label{eq:Br}
\end{eqnarray}

\noindent The errors indicated in Eq.~(\ref{eq:Br}) are the measurement
errors of the parameter $A$, $\rho$, $m_t$ and $Br(K_L^0 \to \gamma \gamma)$
respectively and our theoretical uncertainty.
On Fig.1 we represent the experimental data of the measurement
of $Br(K_L^0 \to \mu^+ \mu^-)$,
%
%
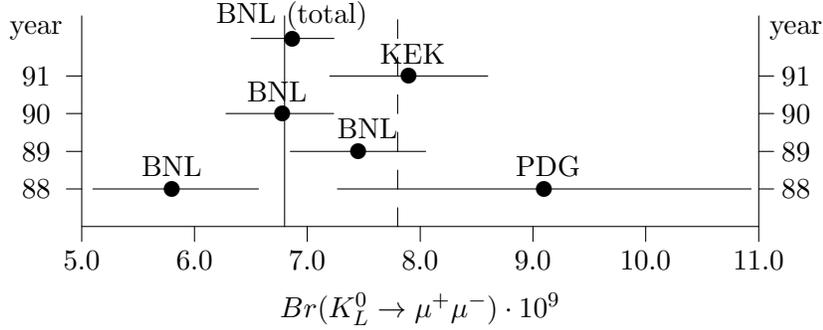
\begin{figure}[tb]
\unitlength=1.00mm
\special{em:linewidth 0.4pt}
\linethickness{0.4pt}
\begin{picture}(110.00,43.00)
\emline{15.00}{13.00}{1}{15.00}{43.00}{2}
\emline{15.00}{15.00}{3}{105.00}{15.00}{4}
\emline{105.00}{13.00}{5}{105.00}{43.00}{6}
\emline{42.00}{15.00}{7}{42.00}{43.00}{8}
\emline{57.00}{15.00}{9}{57.00}{17.50}{10}
\emline{57.00}{20.00}{11}{57.00}{22.50}{12}
\emline{57.00}{25.00}{13}{57.00}{27.50}{14}
\emline{57.00}{30.00}{15}{57.00}{32.50}{16}
\emline{57.00}{35.00}{17}{57.00}{37.50}{18}
\emline{57.00}{40.00}{19}{57.00}{42.50}{20}
\emline{15.00}{20.00}{21}{13.00}{20.00}{22}
\emline{15.00}{25.00}{23}{13.00}{25.00}{24}
\emline{15.00}{30.00}{25}{13.00}{30.00}{26}
\emline{15.00}{35.00}{27}{13.00}{35.00}{28}
\emline{105.00}{20.00}{29}{107.00}{20.00}{30}
\emline{105.00}{25.00}{31}{107.00}{25.00}{32}
\emline{105.00}{30.00}{33}{107.00}{30.00}{34}
\emline{105.00}{35.00}{35}{107.00}{35.00}{36}
\put(27.00,20.00){\circle*{2.00}}
\emline{16.50}{20.00}{37}{38.50}{20.00}{38}
\emline{30.00}{15.00}{39}{30.00}{13.00}{40}
\emline{45.00}{15.00}{41}{45.00}{13.00}{42}
\emline{60.00}{15.00}{43}{60.00}{13.00}{44}
\emline{75.00}{15.00}{45}{75.00}{13.00}{46}
\emline{90.00}{15.00}{47}{90.00}{13.00}{48}
\put(76.50,20.00){\circle*{2.00}}
\emline{49.00}{20.00}{49}{104.00}{20.00}{50}
\put(51.75,25.00){\circle*{2.00}}
\emline{42.75}{25.00}{51}{60.75}{25.00}{52}
\put(41.70,30.00){\circle*{2.00}}
\emline{34.20}{30.00}{53}{48.50}{30.00}{54}
\put(58.50,35.00){\circle*{2.00}}
\emline{48.00}{35.00}{55}{69.00}{35.00}{56}
\put(43.00,40.00){\circle*{2.00}}
\emline{37.55}{40.00}{57}{48.55}{40.00}{58}
\put(9.00,20.00){\makebox(0,0)[cc]{$88$}}
\put(9.00,25.00){\makebox(0,0)[cc]{$89$}}
\put(9.00,30.00){\makebox(0,0)[cc]{$90$}}
\put(9.00,35.00){\makebox(0,0)[cc]{$91$}}
\put(110.00,20.00){\makebox(0,0)[cc]{$88$}}
\put(110.00,25.00){\makebox(0,0)[cc]{$89$}}
\put(110.00,30.00){\makebox(0,0)[cc]{$90$}}
\put(110.00,35.00){\makebox(0,0)[cc]{$91$}}
\put(27.00,23.00){\makebox(0,0)[cc]{BNL}}
\put(77.00,23.00){\makebox(0,0)[cc]{PDG}}
\put(53.00,28.00){\makebox(0,0)[cc]{BNL}}
\put(41.00,33.00){\makebox(0,0)[cc]{BNL}}
\put(59.00,38.00){\makebox(0,0)[cc]{KEK}}
\put(43.00,43.00){\makebox(0,0)[cc]{BNL (total)}}
\put(15.00,10.50){\makebox(0,0)[cc]{$5.0$}}
\put(30.00,10.50){\makebox(0,0)[cc]{$6.0$}}
\put(45.00,10.50){\makebox(0,0)[cc]{$7.0$}}
\put(60.00,10.50){\makebox(0,0)[cc]{$8.0$}}
\put(75.00,10.50){\makebox(0,0)[cc]{$9.0$}}
\put(90.00,10.50){\makebox(0,0)[cc]{$10.0$}}
\put(105.00,10.50){\makebox(0,0)[cc]{$11.0$}}
\put(9.00,41.00){\makebox(0,0)[cc]{year}}
\put(110.00,41.00){\makebox(0,0)[cc]{year}}
\put(60.00,4.00){\makebox(0,0)[cc]{$Br(K_L^0 \to \mu^+ \mu^-) \cdot 10^9$}}
\end{picture}
\caption{Measurements of the branching ratio of $K_L^0 \to \mu^+ \mu^-$
         from BNL E791 and KEK E137 data. The solid line is the unitarity
         limit. The region to the right of the dash line agrees with the SM.}
\end{figure}
%
%
where PDG, BNL and KEK are Particle Data Group,
BNL E791 Collab.~\cite{BNL} and KEK E137 Collab.~\cite{KEK} results.
To the left
of the solid vertical line is the region which contradicts with the unitarity
relation, to the right of the dash vertical line is the region which agrees
with the SM and $B$-physics. As we can see, the KEK result well correlates
with the SM and $B$-physics, whereas the BNL results
are in conflict with the SM. If the tendency of the quest for
$Br(K_L^0 \to \mu^+ \mu^-)$ to the unitarity limit will be verified
by new series of more precise measurements it may be a signal
of a new physics beyond the SM. For example, the real part of the total
amplitude~(\ref{eq:ampl}) can contain an extra term (the contribution
of the relatively light leptoquark~\cite{KM} or something else)
which can cancel sufficiently the contribution of the top quark.


\begin{thebibliography}{15}
%
\bibitem{VSh}
   M.B.Voloshin and E.P.Shabalin, Sov. Phys. JETP Lett. {\bf 23}, 107 (1976).
%
\bibitem{GN}
   C.Q.Geng and J.N.Ng, Phys. Rev. {\bf D41}, 2351 (1990).
%
\bibitem{B}
   L.Bergstrom et al., Phys. Lett. {\bf B249}, 141 (1990).
%
\bibitem{G1}
   A.A.Gvozdev, N.V.Mikheev and L.A.Vassilevskaya, Yad. Fiz.
   (Sov. J. Nucl. Phys.) {\bf 53}, 1682 (1991).
%
\bibitem{G2}
   A.A.Gvozdev, N.V.Mikheev and L.A.Vassilevskaya, Phys. Lett.
   {\bf B274}, 205 (1992).
%
\bibitem{G3}
   A.A.Gvozdev, N.V.Mikheev and L.A.Vassilevskaya, Phys. Lett.
   {\bf B292}, 176 (1992).
%
\bibitem{CDF}
   CDF Collab., F.Abe et al., Preprint FERMILAB-PUB-94/116-E (1994).
%
\bibitem{Ali}
   A.Ali, Preprint CERN-TH.7168/94 (1994).
%
\bibitem{VZSh}
   A.I.Vanstein, V.I.Zakharov and M.A.Shifman, Zh. Eksp. Teor. Fiz.
   {\bf 72}, 1275 (1977).
%
\bibitem{MP}
   E.Ma and A.Pramudita, Phys. Rev. {\bf D22}, 214 (1980); \newline
   T.Inami and C.S.Lim, Progr. Theor. Phys. {\bf 66}, 297 (1981).
%
\bibitem{Ko1}
   P.Ko, Phys. Rev. {\bf D45}, 177 (1992).
%
\bibitem{Ko2}
   P.Ko, Phys. Rev. {\bf D44}, 139 (1991).
%
\bibitem{BNL}
   Y.Kuang (BNL E791 and E871 Collab.), talk presented at the
   {\it 5th Conference on the Intersections of Particle and Nuclear Physics}
   (May 31 -- June 6, 1994).
%
\bibitem{KEK}
   KEK E137 Collab., Akagi et al., Phys. Rev. Lett. {\bf 67}, 2618 (1991).
%
\bibitem{KM}
   A.V.Kuznetsov and N.V.Mikheev, Phys. Lett. {\bf B329}, 295 (1994).
%
\end{thebibliography}
\end{document}